Lori Scott McCabe ✉ ; Jeremiah Williams ; Saikat Chakraborty Thakur ; Uwe Konopka ;
Evdokiya Kostadinova ; Mikhail Pustylnik ; Hubertus Thomas ; Markus Thoma ; Edward Thomas

View Online

Export Citation

### Articles You May Be Interested In

The complete set of Casimir constants of the motion in magnetohydrodynamics

*Phys. Plasmas* (July 2004)

# Experiments and modeling of dust particle heating resulting from changes in polarity switching in the PK-4 microgravity laboratory

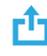

Lori Scott McCabe,[1,2,a)] Jeremiah Williams,[3] Saikat Chakraborty Thakur,[2] Uwe Konopka,[2] Evdokiya Kostadinova,[2] Mikhail Pustylnik,[4] Hubertus Thomas,[4] Markus Thoma,[5] and Edward Thomas[2]

AFFILIATIONS

[1]Physics Department, Mount Holyoke College, 50 College St., South Hadley, Massachusetts 01075, USA
[2]Physics Department, Auburn University, 380 Duncan Dr., Auburn, Alabama 36832, USA
[3]Physics Department, Wittenberg University, Springfield, Ohio 45501, USA
[4]Institut für Materialphysik im Weltraum, Deutsches Zentrum für Luft- und Raumfahrt (DLR), Linder Höhe, 51147 Cologne, Germany
[5]Physikalisches Institut, Justus-Liebig-Universität Giessen, Heinrich-Buff-Ring 16, 35392 Giessen, Germany

[a)]Author to whom correspondence should be addressed: lmccabe@colgate.edu

ABSTRACT

In the presence of gravity, the micrometer-sized charged dust particles in a complex (dusty) plasma are compressed into thin layers. However, under the microgravity conditions of the Plasma Kristall-4 (PK-4) experiment on the International Space Station (ISS), the particles fill the plasma, allowing us to investigate the properties of a three-dimensional multi-particle system. This paper examines the change in the spatial ordering and thermal state of the particle system created when dust particles are stopped by periodic oscillations of the electric field, known as polarity switching, in a dc glow discharge plasma. Data from the ISS are compared against experiments performed using a ground-based reference version of PK-4 and numerical simulations. Initial results show substantive differences in the velocity distribution functions between experiments on the ground and in microgravity. There are also differences in the motion of the dust cloud; in microgravity, there is an expansion of the dust cloud at the application of polarity switching, which is not seen in the ground-based experiments. It is proposed that the dust cloud in microgravity gains thermal energy at the application of polarity switching due to this expansion. Simulation results suggest that this may be due to a modification in the effective screening length at the onset of polarity switching, which allows the dust particles to utilize energy from the potential energy in the configuration of the dust cloud. Experimental measurements and simulations show that an extended time (much greater than the Epstein drag decay) is required to dissipate this energy.

## I. INTRODUCTION

Complex (dusty) plasmas are a four-component plasma system comprised of ions, electrons, and neutral atoms, as well as solid micrometer-sized particles (i.e., "dust"). Dusty plasmas have many fundamental plasma applications such as waves[1–3] and strongly coupled systems.[4–6] Dusty plasmas can also be a diagnostic to the plasma system by using imaging methods and balance of forces to determine properties of the plasma environment.[7,8] A key feature of dusty plasmas is their ability to explore the "atomistic" properties of plasmas and other physical systems at macroscopic scales. This feature will be explored in this work to examine the transformation of configurational energy into thermal energy at the onset of polarity switching under microgravity conditions.

When dusty plasma experiments are performed in ground-based settings, the dust particles become levitated in regions of the plasma where the gravitational force is balanced by the other forces that act upon the particles. Typically, this means that the dust particles become levitated in the plasma sheath, where there can be substantial electrostatic and ion drag forces on the negatively charged dust particles. This leads to dust particles forming thin, flat layers in the vacuum chamber.[9] Therefore, a substantial benefit of performing dusty plasma experiments under microgravity conditions is that the dust particles are no longer restricted to the sheath region of the plasma. The dust particles subsequently can occupy a much larger volume of the background plasma, resulting in the formation of three-dimensional structures.[10] Allowing the dust particles to fill a larger volume beyond just the sheath enables a more detailed study of the forces of dust–dust interactions.

The microgravity experiments discussed in this paper are performed using the Plasmakristall-4 (PK-4) instrument[11] on board the International Space Station (ISS). The PK-4 experiment is a linear, dc glow discharge complex (dusty) plasma device optimized for studying flowing complex plasmas and will be described more completely in Sec. II. For these studies, a constant axial electric field is used to induce a flow of the dust particles, causing the particles to drift into the experimental region (the "flow" segment of the experiment). Once the particles reach the experimental region, a periodic oscillation of the electric field is induced to stop, or "capture," the dust in the experimental region. This periodic electric field oscillation is known as "polarity switching." At the onset of polarity switching, the direct electrostatic and ion drag forces on the dust particles are altered, thereby modifying their dynamics. However, these forces modify the electron and ion transport properties in the background plasma,[12,13] thereby impacting the screening of the dust grains and the subsequent dust–dust interaction forces. The behavior of the dust at the onset of polarity switching is the focus of this work.

In microgravity conditions, the dust–dust interactions have been previously examined by Ivlev et al.[14] During the flow portion of our PK-4 experiment, ion-wakes occur along the electric field "after" the dust grains, similar to Fig. 1(b) of Ivlev et al. At the application of polarity switching, the ions can respond to this polarity frequency on the oscillation frequency timescale, while the dust particle motion is stopped. However, the motion of the ion-wake shifts and occurs on both sides of the dust particles along the axial electric field, as seen in Fig. 1(c) of Ivlev et al. This change in ion-wakes could correspond to a change in the potential energy structure of the system.

Dust particle interactions can be described using a Yukawa-like dust–dust interaction model,[9] seen in the following equation:

$$\Phi_{Dust-Dust}(r) = \frac{Q}{4\pi\epsilon_0 r} e^{-\frac{r}{\lambda_D}}, \qquad (1)$$

where $\Phi$ is the interparticle electric potential, $\epsilon_0$ is the vacuum permittivity constant, $\lambda_D$ is the relevant screening (Debye) length for the system, $Q$ is the dust charge, and $r$ is the interparticle distance. As the screening length increases due to the plasma conditions, the dust–dust interaction potential also becomes larger. At the large screen length limit, the plasma does not screen the dust interactions, and the potential becomes Coulomb-like. If the plasma conditions cause the screening length to grow rapidly enough that the dust cloud cannot respond, this can produce Coulomb explosions.[15] This paper will discuss the behavior of the dust at the application of polarity switching, which suggests a change in the dust's effective screening length as well. This change leads to an increase in the dust's temperature (thermal energy) and a change in the dust cloud structure, and it requires the plasma conditions to settle down for temperature dissipation to occur.

This paper reports on the redistribution of dust particle energy at the application of polarity switching in the PK-4 microgravity experiment and compares these experiments to ground-based data. The experimental observations show that the dust cloud gains thermal energy in microgravity experiments but not in ground-based experiments. These observations are also supported by changes in the cloud morphology, where the dust cloud expands in microgravity but not in the ground-based experiments. Furthermore, the dissipation of this thermal energy persists for a substantially longer period (by an order of magnitude or greater) than the characteristic Epstein drag time scales (collision with neutral particles).[16,17] A mechanism for heating and dissipation is proposed and supported by the results of molecular dynamics simulations. The organization of this paper is as follows: Sec. II will describe the experimental setup of the PK-4 device used for these studies and data processing methods using particle image velocimetry (PIV) techniques. Section III will discuss the experimental results in gravity and microgravity conditions, and Sec. IV will present numeric modeling results that support our experimental findings.

## II. EXPERIMENTAL SETUP
### A. PK-4 instrument

The experiment discussed in this paper uses the Plasmakristall-4 (PK-4) instrument. PK-4 is a multi-user apparatus located in the European Space Agency (ESA) Columbus module of the International Space Station (ISS) that is jointly operated by ESA and the Russian Space Agency, ROSCOSMOS.[11] PK-4 is a U-shaped, glass vacuum chamber with a variety of electrodes that can produce dc or rf plasmas, six different dust shakers to introduce spherical, melamine formaldehyde (MF) particles ranging in size from 1.31 to 10.41 $\mu$m, and a variety of diagnostics systems including cameras, lasers, and a spectrometer. A schematic of the apparatus, adapted from Pustylnik et al., is shown in Fig. 1. The PK-4 instrument on the ISS has identical modules located in France (CADMOS), Germany (DLR), and Russia (JIHT) that are used for ground-based testing and validation. Ground-based results presented in this paper are performed using the PK-4 Science Reference Module located at DLR in Oberpfaffenhofen, Germany.

For the experiments presented in this work, a neon dc glow discharge plasma is generated using the active electrode shown in Fig. 1.

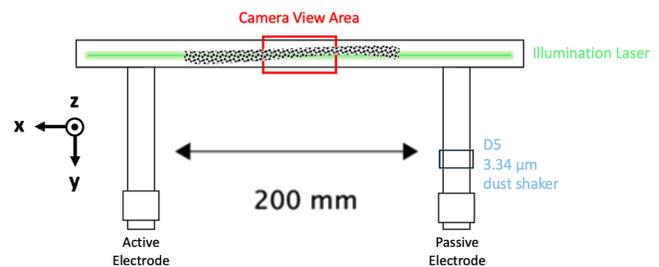

FIG. 1. Schematic of PK-4 experiment on board the International Space Station. The dc current to the active and passive electrodes creates the oscillating electric field for polarity switching. The coordinate system is shown on the left; therefore, the cameras show the x-direction horizontally and z-direction vertically. In both ground-based and microgravity experiments, the x-direction corresponds to the direction of the axial electric field. In ground-based experiments, the vector direction z corresponds to the direction of gravity.

In addition, modulation of the dc current on this electrode is used for polarity switching, to be discussed in greater detail in Sec. II B.

PK-4 experiments are performed using scripts written using a C-scripting language library by the DLR group. Because the PK-4 Science Reference Module instrument and the PK-4 flight instrument on the ISS use the same scripts in both locations, all operating conditions are equal, except gravity and camera height (for sheath/dust-levitation purpose on the ground). These scripts ensure that the timings of the application of polarity switching and experimental conditions (e.g., plasma conditions and electric field) are identical in both locations, allowing for the direct comparison of gravity and microgravity experiments.

PK-4 uses two Basler cameras, denoted as Particle Observation cameras 1 and 2 (PO1 and PO2, respectively), with an overlapping horizontal field-of-view region of $\sim$2 mm. This produces data from a region of $\Delta x \sim 40$ mm. The data presented in this manuscript uses a framerate of 70 fps for both cameras, and each camera has a slightly different scaling conversion ($\mu$m/pixel) in the vertical and horizontal directions, listed in Table IV of Pustylnik et al.[11]

### B. Plasma and dust parameters

The experiments discussed in this paper were performed during Campaign 7, conducted in July 2019. The experiment used a neon dc glow discharge plasma generated with neutral gas pressures, $p = 0.2$–$0.6$ mBar (150–450 mTorr), and discharge currents, $I_{DC} = 0.5$–$1.0$ mA. This paper focuses on measurements made with $p = 0.6$ mBar (450 mTorr) and $I_{DC} = 0.7$ mA to investigate a specific case with certain dust cloud behaviors, while the other cases from this Campaign will be discussed in the future. Following the parameter characterization from empirical models described in Pustylnik et al., the plasma conditions in PK-4 for this specific dataset have the following values: electron density, $n_e = 2.04 \times 10^8$ cm$^{-3}$, electron temperature, $T_e = 8.08$ eV, and an axial electric field, $E_x = 227$ V/m.

For the studies described here, a single particle dispenser (D5) introduces 3.38 $\mu$m diameter MF particles into the experiment (density, $\rho_d = 1510$ kg/m$^3$, and mass, $m_d = 3.05 \times 10^{-14}$ kg). When injected into PK-4, the dust particles generally form an ellipsoid cloud that flows through the field of view of the experiment with peak axial velocities of up to 20 mm/s. During the experiment's flow and capture phases, the interparticle spacing typically varies from 200 to 300 $\mu$m, corresponding to dust number densities $n_d \sim 10^5$ cm$^{-3}$. Representative images showing the flowing and capture phases can be seen in Figs. 2(a) and 2(b), respectively. Using the plasma parameters given above and assuming that the particle charge can be computed using an orbit-motion-limited model,[18] the dust grain charge can be approximated to be $Z_d \sim 6000$ elementary charges. These can be combined to estimate the dust plasma frequency, $\omega_{dp} = 10^3$ rad/s.

The primary technique to capture a dust cloud within the PK-4 apparatus when operating in dc mode is polarity switching, a rapid oscillation of the axial electric field through a periodic modulation of the dc current on the active electrode. For the experiments described in this work, the polarity switching frequency is $f_{ps} = 500$ Hz or $\omega_{ps} = 3140$ rad/s. Compared to the dust plasma frequency, the applied polarity switching frequency is over a factor of 3 higher. This means that the ions and electrons can respond to this oscillation, but the dust particles cannot due to their large inertia. As a result, at the application of polarity switching, the flowing motion of the dust cloud is halted and is "captured." Once the dust cloud is captured, the particles begin to form string-like structures after a few seconds.[14,19–21] The goal of this work is to investigate this initial redistribution of the energy of the dust cloud immediately after (t < 1 s) the application of polarity switching.

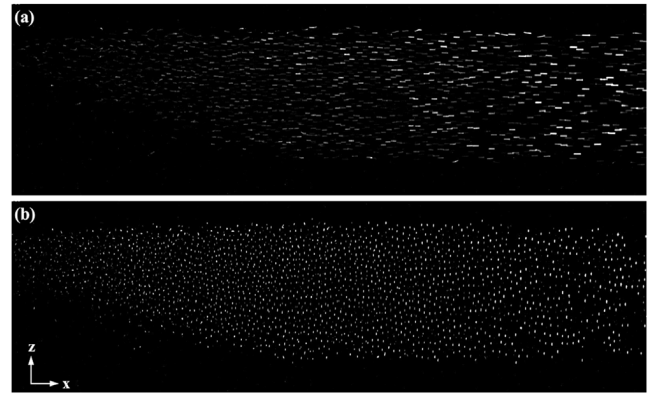

FIG. 2. Black/White images from camera PO1 during the (a) initial injection, "flow" process, and (b) after the onset of polarity switching when the bulk motion of dust has stopped, "capture."

To optimize the limited experimental timeslot, we also utilized a technique called reinjection. This is where the electric field is set to a negative constant value to send the dust cloud out of the field of view in the $-x$-direction, then set to a positive constant value just like an initial injection to bring the cloud back into the field of view, and then repeat the capture using the same oscillation settings for polarity switching as a "standard" injection capture. This allows us to "reset" the cloud and repeat experiments without flushing the system of particles to form a new cloud. We use this reinjection technique to change the plasma discharge current while keeping the same gas pressure. Our analysis indicates that the phenomena associated with the redistribution of the flow kinetic energy occurs for both the initial "injection" and "reinjection" processes.

### C. Imaging and PIV analysis

Particle Image Velocimetry (PIV) techniques are used to analyze the particle motion from the PO cameras. PIV is a particle analysis technique in which an image is decomposed into interrogation cells, containing three or more particles, and cross correlation is performed between image pairs to determine the average velocity vector that corresponds to the group of particles.[22] Dedicated PIV hardware systems can capture images, but image sequences from "high-speed" cameras can also be used with PIV analysis techniques. PIV techniques have been used extensively for dusty plasma studies and earlier work by Thomas et al.,[23] Williams and Thomas,[24] and Fisher and Thomas,[25] specifically for determining the thermodynamic properties of dusty plasmas. Additionally, PIV techniques have been specifically benchmarked against ground-based PK-4 experiments[26] and PK-4 simulations.[27] For the PK-4 measurements reported in this paper, the high-speed imaging technique is used to obtain 2D–2V (two spatial dimensions, two velocity components) vectors in the $x$- and $z$-directions and then used to extract representative thermodynamics quantities.

Because PIV measures the motion of groups of particles, it is particularly well-suited for higher particle number densities and higher speed particle flows. Therefore, PIV is ideally suited for analyzing dust particles in PK-4, as illustrated in the image of flowing particles shown in Fig. 2(a). While Particle Tracking Velocimetry (PTV) techniques do work well for the capture portion of the experiment when drift velocities are low, we seek to measure the velocity space distribution before and after polarity switching, so the PIV technique is used on both portions of the data to obtain self-consistent results.[28] We process each PO camera individually, but we can account for the duplicate vectors in the overlapping region and produce a resultant vector field that encompasses the entire field of view in the PK-4 experiment.

## III. EXPERIMENTAL MEASUREMENTS AND ANALYSIS

PIV techniques are used to characterize the particle motion between two subsequent video frames and generate a vector field, Fig. 3(a), that quantifies the particle motion in the field of view of the camera. These vector fields are used to generate velocity distributions in each vector direction and fit using a Maxwell–Boltzmann distribution, $f(v) = \sqrt{\frac{m}{2k_B T}} e^{-\frac{m(v-v_{drift})^2}{2k_B T}}$, to extract the drift velocity, $v_{drift}$, and the kinetic temperature of the dust, $T$. Representative distributions are seen in Figs. 3(b) and 3(c) for a single camera (PO2). This technique is applied before and after the onset of polarity switching to extract the time evolution of the thermal state of the system. This is done for the ground and microgravity experiments, described in Sec. II. We begin the discussion of the experimental observations by describing the observations made during the ground-based studies and then discussing how those results motivated the microgravity studies.

An example of the time evolution of the drift velocity and effective dust kinetic temperature of the dust is shown in Fig. 4. In this case, the operating pressure is $p = 0.6$ mBar, the dc discharge current is set at $I_{DC} = 0.7$ mA, and the polarity switching frequency is 500 Hz. For all the data shown for both the ground experiments (and later for the microgravity experiments), polarity switching is defined and shifted to occur at $t = 0$ s. As noted previously, the camera frame rate for all these experiments is 70 frames per second, corresponding to a $\Delta t = 0.014$ s interval between each measured data point. Figure 4 shows the extracted drift velocity and effective kinetic dust temperature from fits to the measured velocity distributions in the $x$- (i.e., parallel with the axial electric field) and $z$- (i.e., transverse to the axial electric field) directions as a function of time for both the gravity [Fig. 4(a)] and microgravity [Figs. 4(b) and 4(c)] experiments from PO1, and gravity [Fig. 4(d)] and microgravity [Figs. 4(e) and 4(f)] experiments for PO2.

Starting with the ground-based observations in Figs. 4(a) and 4(b), it is observed that within 1–2 video frames after polarity switching (i.e., $\Delta t \leq 0.028$ s), there is a decrease in the velocity of the particles from their drift speed from $v_{drift} \sim 10$–15 mm/s to $v_{drift} < 1$ mm/s, Figs. 4(a) and 4(b) top. The timescale for this reduction in the drift velocity is generally consistent with the slowing of the dust particles due to collisions with the neutral atoms, i.e., Epstein drag,[17,29] where the dust–neutral collision frequency is estimated to be $f_{epstein} = 88.3$ s$^{-1}$. This corresponds to a damping time of $\frac{1}{f_{epstein}} \sim 0.011$ s, which is equivalent to $\sim 1$ video frame and consistent with the experimental observations.

For the gravity-based measurements, the presence of waves in the lower part of the cloud is observed. Because our goal is to investigate the conditions of the flowing particles, the PIV analysis configuration was optimized for the flow and capture phases of PK-4 and not for the waves, as illustrated in Fig. 4. Therefore, when the waves are present, we have less accurate flow and temperature measurements before t = 0 s, and this data is overlaid with gray boxes in Figs. 4(a) and 4(d). In the gravity-based x-direction effective temperature [Fig. 4(a), bottom], there appears to be a momentary increase in temperature, and then the temperature quickly drops back to ambient dust cloud temperatures. These results are consistent with our preliminary ground-based PK-4 work and motivated our work to determine whether microgravity conditions would allow us to reveal additional details of the apparent change in the thermal properties of the dust cloud that may be occurring at the onset of polarity switching.

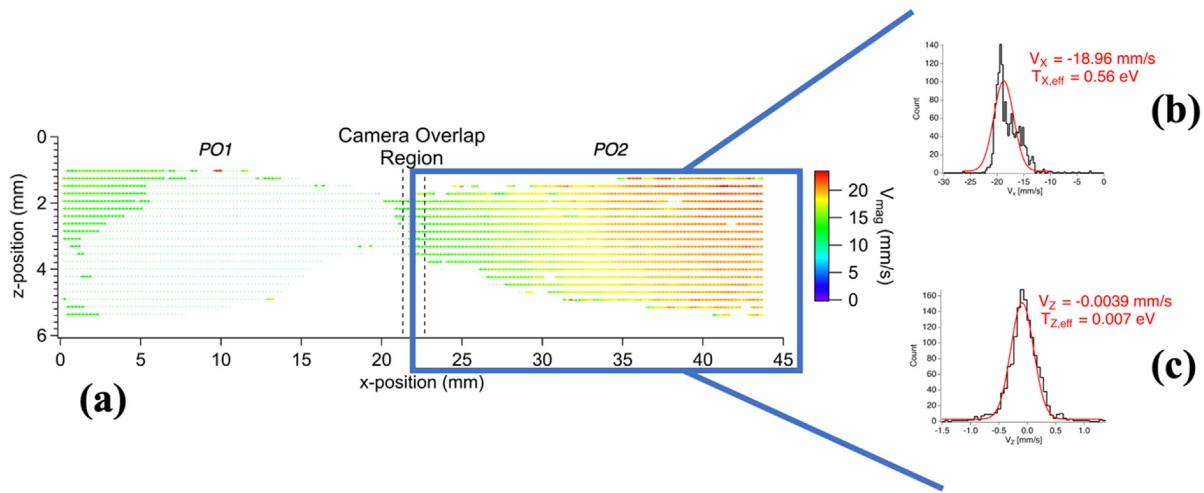

FIG. 3. (a) A representative velocity vector field from both cameras with p = 0.6 mBar, I = 0.7 mA, with PO1 having the field of view x = 0–23 mm, and PO2 x = 21–44 mm. This dust cloud was split into two smaller clouds at capture, so fewer vectors (and smaller magnitudes) from x = 10–20 mm. Distributions of the measured velocity from the data of PO2 in the (b) x- and (c) z-directions.

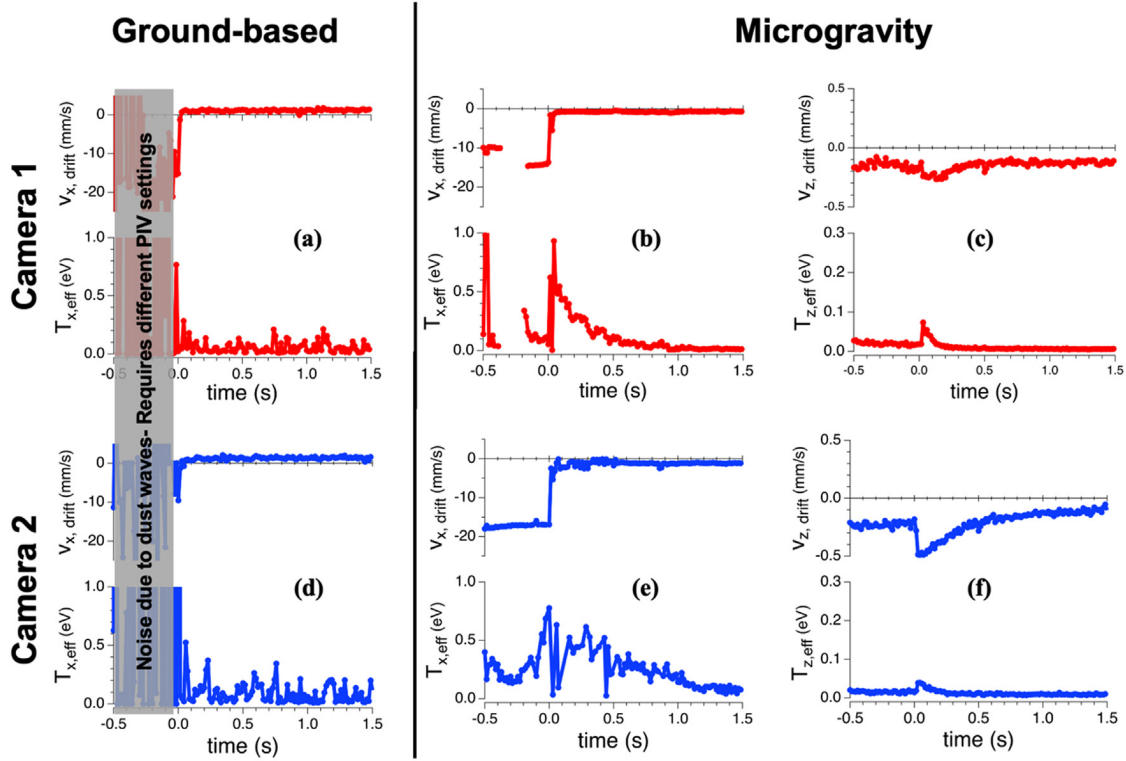

**FIG. 4.** Plots of the drift velocity and effective dust kinetic temperature as a function of time from (a) and (d) ground and (b) and (c), (e) and (f) microgravity experiments using data from (red) PO1 and (blue) PO2 cameras in the (a), (b), (d), and (e) x- and (c) and (f) z-directions. The onset of polarity switching occurs at t = 0 s. There is heating in the dust cloud at the application of polarity switching observed in both ground and microgravity experiments. In contrast, an extended cooling time is only observed under microgravity conditions.

Figures 4(b) and 4(e) and 4(c) and 4(f) show measurements of the x- (parallel to the axial electric field) and z- (transverse to the axial electric field) components of the particle response to polarity switching for an experiment performed under microgravity conditions on the ISS. It is first noted in Figs. 4(b) and 4(e) that there is a decrease in the horizontal (x-component) drift velocity of the particles that nearly precisely matches the ground-based experiments in Figs. 4(a) and 4(d). Since both experiments are performed under the same gas pressure and discharge current conditions, we can conclude that, with respect to the flow, in both the ground-based and microgravity experiments, the particle drift decays in 1–2 video frames on a timescale that is consistent with Epstein drag. However, in terms of the thermal response of the system, there is a substantial difference between the ground and microgravity systems.

The effective temperature in the x-direction under microgravity conditions, Figs. 4(b) and 4(e) lower graph, shows a significant rise in the temperature on the timescale of 1–2 video frames, followed by an extended cooling time that is significantly more than what is expected based on Epstein drag. A careful examination of the PIV settings was performed to ensure that this was not an experimental or analytical artifact, and several analysis approaches are discussed below to confirm the validity of this observation. This observed heating and dissipation occurs in multiple capture datasets at various combinations of plasma operating conditions (i.e., pressure and current); we are just choosing to illustrate this case as it is the most pronounced result. The microgravity z-direction effective temperature, Figs. 4(c) and 4(f) bottom, also shows a minor increase in the temperature and an extended cooling event, which may indicate an additional dissipation of the thermal energy into both the parallel and perpendicular directions (relative to the flow) at the application of polarity switching. These responses vary slightly from the two cameras. Still, the response trends are present throughout the dust cloud, indicating a more robust interaction occurring at the application of polarity switching.

While the trend in the two cameras' data is the same in the microgravity conditions, we want to confirm why PO1 has much more pronounced changes in the effective temperature and drift velocity compared to PO2. Figure 3 shows the resultant vector field when both cameras are processed through PIV and combined appropriately. The region x = 0–22 mm is from camera 1 and x = 22–44 mm is from camera 2. This illustrated experiment had a split in the cloud, so few vectors were returned in the region x = 10–20 mm. It is important to note the varying magnitude of velocity in the right half of the cloud. This magnitude variance indicates a spatially non-uniform response to the application of polarity switching. To better account for the variations in the dust cloud distribution functions, we repeated our analysis processes using a sub-divided regions approach to better determine the energy dissipation throughout the dust cloud at the application of polarity switching.

A second approach to investigating the cause of the rise in effective temperature at the application of polarity switching is to identify the spatial non-uniformity by dividing the frames into four regions and repeating the analysis techniques, as described in Fig. 3. The red data shown in Fig. 5 is the original PO1 from Fig. 3(b), and the four other datasets correspond to a specific subregion, as indicated by the dust cloud frame directly above. By comparing the individual regions and whole field datasets, the rise in temperature comes from only part of the cloud, in this case, most strongly from regions 1–3. This further indicates a non-uniform response in the dust cloud at the application of polarity switching and confirms that the heating at the onset of polarity switching throughout the cloud is real in the experiment and not an analysis artifact.

To further investigate the differences in the microgravity cloud morphology compared to the ground-based data, we can investigate the interparticle spacing of the dust under microgravity conditions after the onset of polarity switching, shown in Fig. 6. We find all particles' positions in the dust cloud using particle tracking and then calculate the pair correlation function[30,31] to get the interparticle spacing per video frame. This analysis is done using the TrackPy package,[32] which has sub-pixel resolution capabilities. After applying polarity switching, the interparticle spacing increases slightly, ~1 pixel. This expansion occurs on the order of 0.5 s, which is the same timescale as the increase in perpendicular drift velocity shown in Fig. 4(f) and the decay of the effective temperature seen in Fig. 4(b). The interparticle spacing and momentary drift velocity change combine to support an expansion of the microgravity dust cloud. We repeated this analysis for ground-based experiments but saw no such morphological change in the dust cloud.

Work done by Hartmann et al.[12] and Matthews et al.[33] using high-speed measurements in the ground-based PK-4 setup at Baylor University (PK-4 BU) reveal that changes in the dc electric field in PK-4 lead to short, $\mu$s-scale modifications of the plasma as detected by changes in the propagation of ionization waves.[34,35] Based upon these experimental observations from the Baylor group and a wide range of simulation tests (presented below), we have developed the following hypothesis for the observed dust heating in the microgravity cloud. As the plasma momentarily changes/collapses when there is a modification in the dc electric field, the dust particles will lose a small amount of charge, but the dust will be able to interact with a significantly larger number of dust particles due to a reduction in the screening by the plasma. This results in the dust cloud undergoing a Coulomb explosion,[36,37] and some of the energy stored in the configuration of the dust (configurational/potential energy) is transformed into thermal energy, resulting in the observed heating event. This collapse happens on a timescale that is an order too fast for the frame rate of the camera on the PK-4 instrument to observe but has been observed elsewhere.[12,34] As the plasma reforms under the new, oscillating conditions, the charge on the dust returns to its initial value, and the dust continues to expand while the plasma screening length decreases back to the nominal values described in Sec. II B. During this process, the dust continues to expand, while a small fraction of the configurational energy of the dusty plasma continues to be converted into thermal energy. This results in the extended cooling times that are observed in the experiments.

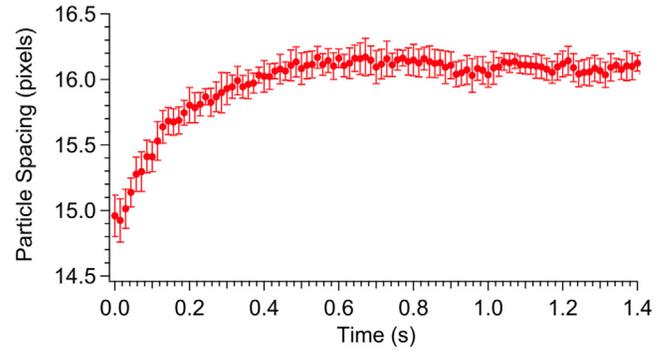

FIG. 6. Measured interparticle spacing as a function of time after the application of polarity switching at t = 0 s in microgravity. This increase is in the same order as the effective temperature decay shown in Fig. 3.

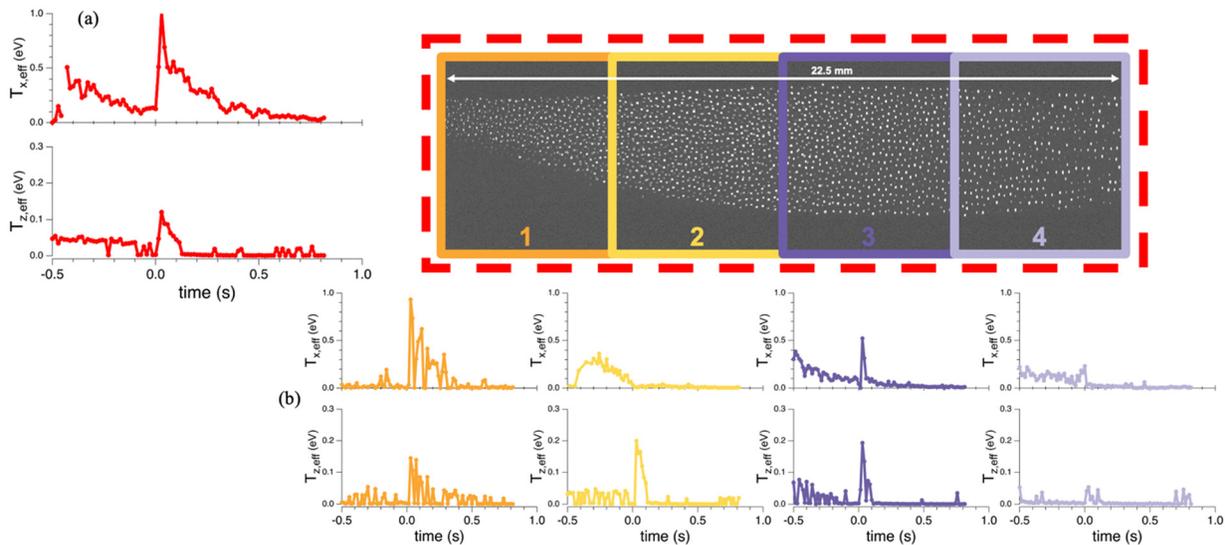

FIG. 5. (a) The same effective dust cloud temperature data from PO1 presented in Figs. 3(b) and 3(c), reorganized to have Tx on the top and Tz on the bottom. (b) A sample frame divided into four regions, and the boxes marking the regions match the color to the region's temperature data directly below. All four regions show a rise in temperature at the application of polarity switching, but there is a non-uniform response in the magnitude of the temperature increase. This suggests a non-uniform response throughout the cloud to the application of polarity switching. Note the difference in the temperature axes scales between Tx and Tz.

## IV. NUMERICAL MODELING

To properly understand the physical processes that may be contributing to the heating, we use a numerical model to replicate and gain more insight into this experiment. We utilize a molecular dynamics code, YOAKµM (Yukawa-Ordered and Kristallized Microparticle model),[38] created by the Thomas group in 2018 utilizing C++ language. For the context of this paper, YOAKµM uses a fourth-order Runge–Kutta algorithm to solve either two- or three-dimensional particle dynamics for up to several hundred charged dust particles in a background plasma. While the particle dynamics are evaluated self-consistently, the charge on the particles is calculated from the plasma parameters and is held at a fixed value for computational efficiency. The particle interactions are governed via interparticle forces (Coulomb and Yukawa-screened), constant and oscillating electric fields, neutral drag, and a thermal (Langevin/Brownian motion) heater set at room temperature. These forces allow for the recreation of the experimental conditions of PK-4 using the system of equations,

$$m_i \frac{d\vec{v}_i}{dt} = q_i \nabla \Phi_{Dust-Dust} + \vec{F}_E + \vec{F}_{drag} + \vec{F}_{Thermal}, \quad (2)$$

where the first term of the interparticle electric potential, $\Phi$, can be expanded to

$$q_i \nabla \Phi_{dust-dust} = \frac{q_i Q_j}{4\pi\varepsilon_0} e^{-\frac{r_{ij}}{\lambda_{screening}}} \left( \frac{1}{r_{ij}^2} + \frac{1}{r_{ij} \lambda_{screening}} \right), \quad (3)$$

where q and Q are the dust particle charges, r is the interparticle distance between particles, and $\lambda$ is the effective screening length of the dust–plasma system. During the flowing part of the simulation, this effective screening length for the dust particles will be considered equivalent to the electron Debye length, $\lambda_{De}$. After the application of polarity switching, variations in the screening length will be done manually with different segments of code and will be used to represent the time evolution of all the changing dust–plasma interaction conditions after the application of polarity switching[12] that will be described in further detail below.

To simulate a dust cloud in PK-4, a cloud of ~1000 particles was initialized in a random orientation (position, velocity) and is allowed to evolve according to Eq. (2) to create the stable dusty plasma crystal in thermal equilibrium using a thermal heater and neutral drag force at the respective calculated magnitudes of the experimental conditions listed in Sec. II B based on pressure and current. The cloud is subsequently "injected" into the simulation by applying an electric field at the same magnitude in PK-4 based on the empirical model in Table II by Pustylnik et al.[11] It is noted that the magnitude of the drag coefficient is adjusted in the code to match the magnitude of the particle drift velocity measured during the experimental injection process. We then simulate polarity switching by replacing the constant electric field with an oscillating field at a frequency of 500 Hz, as is done in the experiment. Because the positions and velocities of the particles are recorded at each time step in the simulation, it is then possible to construct a velocity distribution and extract an effective kinetic temperature of the dust at each simulation time step.

We replicate our microgravity experimental results using YOAKµM through careful consideration of the simulation steps. Our initial simulation step uses dust-scale time steps ($\Delta t_{step} \sim 1$ ms) to model the flow state of the experiment. However, the underlying processes that govern the heating mechanism operate at a much smaller timescale, so we then reduce the simulation time steps to 1 µs before introducing the changing screening lengths to capture the entire physics of the following results. We have incorporated this phenomenon into the YOAKµM simulation in the following manner: at the onset of polarity switching, a 2.5 ms suppression of the plasma[12,13] is modeled by allowing the screening length of the particles to become disturbed, i.e., $\lambda_{screening}$ is allowed to become large, and the particles are allowed to have a more Coulomb-like particle interaction during this period. This 2.5 ms is empirically chosen to best reproduce the heating event numerically. This leads to a Coulomb expansion of the cloud as electrostatic potential energy between the charged dust particles is converted into kinetic energy. The plasma is restored at the end of this 2.5 ms period with a new effective screening length between particles. If the screening length is too large (i.e., Coulomb-like), the cloud continues to Coulomb expand, and if it is too small, it rapidly decays into an ordered dust cloud.

Figure 7(a) shows the results of several simulations. The simulation results are shown in the varying pink-colored curves, and the experimental measurements from PK-4 are shown as the gray data points. All the simulations follow the process previously described through the step with a Coulombic particle interaction for 2.5 ms, the light gray curve, to model the creation, injection, and heating event for the dust cloud, and the cloud is observed to expand during the 2.5 ms during which the Coulomb explosion occurs in the simulation. The position and velocity information of the particles are then used as the input conditions for the final step of the simulation, where effective screening lengths, $\lambda_{screening}$, are varied from 1 to 3 mm. For reference, the electron Debye length for these conditions based on the parameters in Pustylnik et al.[11] is $\lambda_{De} = 1.45$ mm.

The results of the simulations suggest that immediately after the heating event, an effective screening length $\lambda_{screening} \sim 3$ mm for the first video frame, followed by an effective screening length of ~2 mm, provides an effective fit to the experimental observations, and these are both larger than the pre-polarity switching electron Debye length. However, this agreement only persists from $0.0 < t < 0.25$ s. For $0.25 < t < 0.4$ s, an effective screening length of $\lambda_{screening} \sim 1.5$ mm appears to be in reasonable agreement with the experimental measurements. Then, for $t > 0.4$ s, $\lambda_{screening} \sim 1.0$ mm provides reasonable agreement with the experimental data. This change in dust screening length vs time is visualized in Fig. 7(b). To further help coordinate between the two figures, the x-axis ticks in the screening length time-variation figure are at the framerate of the experiment to help match when the simulations and the experimental effective temperature values intersect to the exact data point. In the equation of motion, Eq. (2), this extended decay passing through the various screening lengths would suggest that the dominating term influencing this extended decay timescale is the interaction force, which is expanded in Eq. (3).

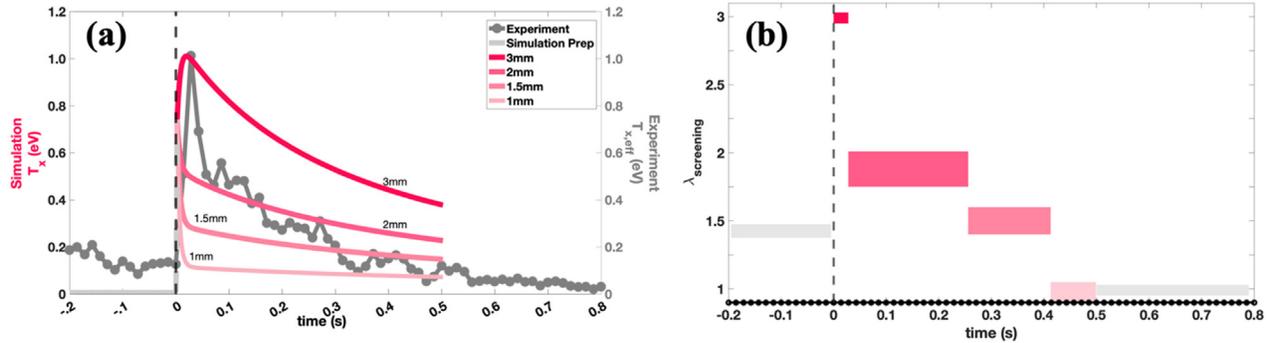

**FIG. 7.** Plot of (a) the effective dust temperature in the x-direction from (dots) experiment [same as Figs. 3 and 5(a)] and (solid curves) simulation for different dust screening lengths, and (b) estimated effective dust screening lengths as a function of time after the application of polarity switching at t = 0 s. A 3 mm screening length provides the best fit to the experimental data briefly but then quickly overestimates the temperature, indicating smaller screening lengths are needed to match the experimental measurements at later times.

The results shown here suggest that the MD simulations performed using the YOAK$\mu$M code provide a reasonable fit to the experimental observations. These results also suggest an effective larger screening length than expected immediately following the application of polarity switching on the experimental conditions. On the assumption that the screening length is representative of an effective Debye length and that its variation is dominated by a change in the plasma density,[12,13] the decreasing screening length observed in the simulations would suggest an increasing plasma density for a short period after the application of the polarity switching; this would be consistent with the modification in the plasma shown by Matthews *et al.*[33]

## V. SUMMARY

A large increase in effective dust temperature in the x-direction has been observed under microgravity conditions. It has been identified as a distinct effect when comparing data from ground and microgravity experiments. After a detailed investigation, it was established that the dust cloud heating and subsequent extended temperature decay were not due to any artificial results but an accurate response to a change in the plasma conditions. Further investigation also revealed an asymmetric response throughout the dust cloud at the application of polarity switching. This unique microgravity response is also visually supported by the dust cloud expanding at the application of polarity switching under microgravity but with no morphological change in the ground-based experiments.

Based on our simulations, we can replicate the thermal response to the change in the dust–plasma system at the application of polarity switching in microgravity experiments. Since the thermal response depends on the screening length in our numeric simulations, we hypothesize that when the plasma changes parameters, it causes a brief structure collapse in the dust cloud, and the screening length decreases. The screening length returns to the original value as the plasma returns to a new dynamic equilibrium, but since the dust cloud collapsed during this small timescale plasma change, the cloud experienced a Coulomb-like explosion, resulting in a net expansion in the dust cloud size. By changing the dust screening length in YOAK$\mu$M, the system can self-regulate the decay of the temperature, which is on the same timescale that occurs in our microgravity experiments. The change in screening length likely arises from a cyclic response of a combination of the temperature and density of the plasma system changing and, therefore, a momentary drop in accumulated charge on the dust particles.

In the microgravity experiment, a heating event occurs at the time of capturing the dust particle by polarity switching. This heating is also reproducible using an MD simulation when using a larger effective screening length than a calculated Debye length would suggest for the plasma conditions. We suspect that this larger screening length is caused by a momentary collapse of the plasma system as the electric field settings change from uniform to oscillating. This heating occurs in other changes of the polarity switching settings as well, and future work will focus on the different plasma conditions' capacity to allow the dust cloud to gain thermal energy.

## ACKNOWLEDGMENTS

Funding for this work is provided by the NASA/Jet Propulsion Laboratory through Grant No. JPL-RSA 1571699. Support for authors L.S.M., E.T., and S.C.K. is provided by the National Science Foundation EPSCoR Project, Connecting the Plasma Universe to Plasma Technology in Alabama (CPU2AL), through Grant No. OIA-1655280. The research activities of JW are also supported in part by the U.S. National Science Foundation through its employee IR/D program. Support for author E.K. is provided by the National Science Foundation through Grant No. NSF-PHY-2308742. M. Toma is supported by the German Aerospace Center (DLR) under Grant No. 50WM2044. The PK-4 instrument is the joint ESA-ROSCOSMOS project on-board the ISS under Contract No. DLR/BMWi 50WM1441. The authors acknowledge A. Lipaev, A. Zobnin, and A. Usachev at the Joint Institute for High Temperatures for their assistance in executing the experiment and support of this article.

## AUTHOR DECLARATIONS
### Conflict of Interest

The authors have no conflicts to disclose.

## Author Contributions

**Lori Scott McCabe:** Conceptualization (equal); Data curation (equal); Formal analysis (equal); Investigation (equal); Methodology (equal); Visualization (equal); Writing – original draft (equal); Writing – review & editing (equal). **Jeremiah Williams:** Conceptualization (equal); Data curation (equal); Funding acquisition (equal); Investigation (equal); Methodology (equal); Visualization (equal); Writing – review & editing (equal). **Saikat Chakraborty Thakur:** Formal analysis (equal); Investigation (equal); Writing – review & editing (equal). **Uwe Konopka:** Conceptualization (equal); Funding acquisition (equal); Methodology (equal). **Evdokiya Kostadinova:** Conceptualization (equal); Investigation (equal); Methodology (equal). **Mikhail Pustylnik:** Data curation (equal); Methodology (equal); Software (equal); Writing – review & editing (equal). **Hubertus Thomas:** Funding acquisition (equal); Project administration (equal); Writing – review & editing (equal). **Markus Thoma:** Funding acquisition (equal); Project administration (equal); Writing – review & editing (equal). **Edward Thomas:** Conceptualization (equal); Funding acquisition (equal); Methodology (equal); Project administration (equal); Writing – review & editing (equal).

## DATA AVAILABILITY

The data that support the findings of this study are available from the corresponding author upon reasonable request.